\documentstyle[12pt]{article}
\begin{document}
\title{Coherence of Currents in Mesoscopic Cylinders}

\author{M. Stebelski, M. Szopa and E. Zipper\\
 Institute of Physics, University of Silesia,\\
 Uniwersytecka 4, 40-007 Katowice, Poland}

\baselineskip=15pt

\date{\today}
\maketitle

\begin{abstract}
   The persistent currents driven by the pure Aharonov-Bohm type magnetic field
in mesoscopic normal metal or semiconducting cylinders are studied. A 
two-dimensional (2D) Fermi surfaces are characterized by four parameters. 
Several conditions for the coherence and enhancement of currents are discussed.
These results are then generalized to a three-dimensional (3D) thin-walled 
cylinder to show that under certain geometric conditions on the Fermi surface, a 
novel effect - the appearance of spontaneous currents is predicted.\\ 
PACS numbers: 71.30.+h, 72.10.-d\\
Keywords: mesoscopic cylinder, Fermi surface, spontaneous currents.

\end{abstract}

\section{Introduction}

Quantum coherence in mesoscopic systems arose a great interest in the last
years because it leads to many interesting phenomena such as persistent
currents, conductance oscillations etc. \cite{ww}. In a series of papers
\cite{Sz,szz} we investigated persistent currents in a collection of
isolated mesoscopic rings stacked along one axis. The current $I(\phi)$
induced by the magnetic flux $\phi$, due to the Aharonov-Bohm effect, was 
therein calculated under the
assumption, that the electrons from different rings interact via the
magnetostatic (current-current) interaction and the interaction was
considered in the mean field approximation.  It was shown that due to the
selfconsistent equation for the current, in some cases, a novel effect -
the spontaneous selfsustaining currents at the zero external flux - can be
predicted. A similar effect, due to the Aharonov-Casher effect was also 
recently proposed \cite{choi}.

In the present paper we investigate persistent currents in 2D and 3D
cylinders made of a clean normal metal or a semiconductor (ballistic
regime).  It is known that the magnitude of the currents in 2D systems
depends strongly on the correlations of currents from different channels
\cite{cheu}. To study this correlation we calculate the currents in
systems with different shapes of the Fermi surface (FS). We also search
for the possibility of obtaining spontaneous currents for the most
favorable FS shapes.

\section{Persistent Currents in a Single Channel}
   
   Let us consider a 2D normal metal or semiconductor system
of a cylindrical geometry.  The circumference and height of the cylinder
are denoted by $L_{x}$ and $L_{y}$ respectively. We neglect the width of
its wall, assuming it is of the order of $\lambda_{F}$ - the Fermi
wavelength. We also assume that the magnetic flux $\phi$ threads the
cylinder axially and is confined to its center, so that the electrons move
in a field-free space. Here we use the cylindrical coordinate system, in
which the height coordinate is y and the azimuthal angle is $\Theta \in
[0; 2 \pi)$. It is convenient to replace $\Theta$ by $x = \frac{L_{x}
\Theta}{2 \pi}$ so that x varies between 0 and $L_{x}$. 

   For the wave function of an electron in the cylinder we apply periodic
boundary conditions in the azimuthal direction and infinite potential
barrier along the y axis. In presence of the magnetic field $\phi$ the wave
function obeys the following boundary conditions \cite{By}
    \begin{eqnarray}
    \label{warbrzeg}
    \psi (L_{x}) = \exp \, \left(\frac{2\pi i\phi}{\phi_{0}}\right)\: 
    \psi (0), \nonumber \\
     \\   
    \left. \frac{d\psi}{dx} \right|_{x = L_{x}} = \exp \, \left( \frac{2 \pi 
    i\phi}{\phi _{0}} \right) \: \left. \frac{d \psi}{dx} 
    \right|_{x = 0} , \nonumber
    \end{eqnarray}
where $\phi_{0} = \frac {h}{e}$ denotes the flux quantum.

    Solving the Schr\"{o}dinger equation in the free electron 
approximation we get quantized energy levels
    \begin{equation}
    E_{qr}(\phi)= \frac{\hbar ^{2}}{2m} \, [k_{x}^{2}(q,\phi) + k_{y}^{2} (r)],
    \label{energy}
    \end{equation}
where $k_{x}(q,\phi)$ and $k_{y}(r)$ are components of the wave vector
$\bf{k}$ in the azimuthal and y direction, respectively, 
  
    \begin{equation}
    k_{x}(q,\phi)= \frac{2\pi}{L_{x}}(q+ \frac{\phi}{\phi _{0}}), \; \;
    \; \; q = 0, \pm 1, \pm 2, \ldots, 
    \end{equation} 
    
    \begin{equation}
    k_{y}(r)= \frac{r\pi}{L_{y}},\; \; \; \; r=0,1,2,\ldots.
    \label{ky}
    \end{equation}

In our 2D system we are interested in the net current along
the azimuthal direction. The Fermi surface (FS) is in this case
2D.  Apart from momentum in the $k_{x}$ direction the
electrons can move in the y direction. The collection of them, having the
same $k_{y}$ or $r$ quantum number will be called a channel. Each channel
is a semi one-dimensional (1D) system with a characteristic radius $K_{x}$,
which is the maximal possible value of the $x$ component of the wave
vector $ \left| k_{x} \right| \leq K_x$ . This radius as well as the
number of channels $M$ depend on the shape of the 2D FS. 

The current carried by a single electron of the energy
$E_{\alpha}$ is given by
    \begin{equation}
    I_{\alpha}(\phi)=-\frac{\partial E_{\alpha}(\phi)}{\partial \phi},
    \end{equation}
where $\alpha$ is an arbitrary set of indices. In case of the energy 
(\ref{energy}) it is a linear function of $\phi$.
The total current in the 2D system is the sum over the
contributions $I_{\alpha}$ of all states, where the occupation probability 
is the Fermi-Dirac (FD) function.
    
In the present and in the the next chapter we assume that the
temperature is zero. We first recall the current $I_{r}(\phi)$ 
corresponding to a single channel $r$ 
with N electrons. It is a sum over $q$ of currents $I_{qr}(\phi)$ 
corresponding to the first
N lowest lying states $E_{qr}(\phi)$. In the following we use the Fourier 
series expansion formula \cite{cheu, Ho} for this current 

    \begin{equation}
    \label{ifi}
    I_{r}(\phi)=\sum_{l=1}^{\infty} \frac{2I_{1}}{\pi l} \cos (lK_{x}L_{x})
    \sin \left( \frac{2\pi l \phi}{\phi_{0}} \right),
    \label{I1D}
    \end{equation}
where 
    
    \begin{equation}
    \label{izero}
    I_{1}=\frac{ev_{F}}{L_{x}} = \frac{e\hbar K_{x}}{mL_{x}}
    \end{equation} 
is the amplitude of the current in the
1D system and $ K_{x}=\frac{\pi N}{L_{x}}$ is the
1D radius of the FS corresponding to the channel r ($v_{F}$
is the velocity of electrons at the FS). The total current for the single
channel $r$ has a sawtooth shape with the period 1 in
$\frac{\phi}{\phi_0}$ and the amplitude inversely proportional to the
circumference of the ring. The current for a system with an even number of
electrons has a discontinuous jump up at $\phi = 0$ whereas in the case of
an odd number of electrons at $\frac{\phi}{\phi_{0}} = 0.5$.  More
generally if the number of electrons in the system varies between even and
odd as a function of $\phi$, the current assumes one of the above values
corresponding to the number of electrons at the particular value of the
flux \cite{cheu,szo1}. In the next chapter we show that this property to
some extent "survives" the transition to a 2D system.

\section{The Currents Depending on the Shape of the Fermi Surface}   
      
The total current in the 2D case  
depends on the strength of the interchannel
correlations. It has been already shown \cite{cheu} that if the
FS is spherical then, for short cylinders $L_{x} \geq L_{y}$ the
channel currents add without phase correlation, whereas for long
cylinders there exist some phase correlation and the total
current is bigger. A large phase correlation among channel
currents means that the increase of the flux $\phi$ results in
an almost simultaneous cross of the FS by the large number of
channels. The most favorable situation takes place if the
separation between the last occupied level and the FS from
channel to channel is nearly the same. There exists then a
perfect correlation among the channel currents because the $M$
levels cross the FS simultaneously while the flux is changed by
one fluxoid - we get then the largest amplitude of the total
current. 

The phase correlation and the value of the total current depend on the
shape of the FS. In case of the free electron approximation considered in
chapter 2 the FS is circular
  
    \begin{equation}
    \label{sfs}
    K_{F}^{2}=K_{x}^{2}+K_{y}^{2}.
    \end{equation}
It is characterized by a single parameter $K_{F}$, specifying its volume. 
The radius $K_{x}$ of the FS in the given channel $r$  
can be easily calculated from (\ref{sfs}).

However for the interacting electrons moving in the crystal lattice the
dispersion relation $E(\bf{k})$ is in general a complicated function of
the wave vector $\bf{k}$, leading to different shapes of the FS. For
example, in the tight binding approximation, for 3D systems we can get,
depending on the crystal symmetry and on the filling factor, the FS being
the sphere, the cube, the cube with rounded corners, octahedron etc.
\cite{anse}. The FS can be in addition differently oriented with respect
to $k_{\alpha}$ ($\alpha = x, y, z$) axes in the reciprocal lattice. For 2D
systems or for 3D systems with 2D conduction (an example of such systems
are high $T_{c}$ superconductors in a normal state) the FS can change from
a circle to a square \cite{ki, crac} via different other shapes. 

   In the following we study the influence of the shape of
the FS on persistent currents in multichannel systems changing
its shape in the following way 
  
    \begin{equation}
    \label{dwym}
    K_{F}^{n}=\left( \left| \frac{K_{x}}{\alpha} \right| \right)^{n} + \left(
    \frac{K_{y}}{\beta}\right)^{n}.
    \label{fersurf}
    \end{equation} 
The shape of this Fermi surface depends on four
positive parameters:  $K_{F}$, n, $\alpha$ and $\beta$. Generally speaking
the parameter n controls the convexity of the FS, whereas $\alpha$ and
$\beta$ measure its curvature at the places where it crosses $k_x$
or $k_y$ axes. To understand how the above parameters influence FS and the
corresponding current in this chapter we consider three cases
in which some of the parameters are kept constant while the other are
varied.

  \begin{description}

   \item[(i)] $K_F = const.$, $n=2$ and $\alpha$, $\beta$ are varied:
    \label{jeden}
  
    \begin{equation}
    \label{elip}
    K_{F}^{2}=\frac{K_{x}^{2}}{\alpha^{2}}+\frac{K_{y}^{2}}{\beta^{2}}.
    \label{elliptical}
    \end{equation}
    
    In this case the FS is elliptical with its major axis  
    along $k_{x}$ or $k_{y}$ depending on whether $\alpha>\beta$ or $\beta>
    \alpha$, respectively. The curvature of the FS at the point, where it 
    crosses the $k_{x}$ axis is equal to $\frac{\beta}{\alpha} K_F$. We 
    assume in addition that  $\alpha \beta =1$ 
    because it ensures the independence of the volume under the FS on 
    $\alpha$ 
    and $\beta$. Under this assumption the number of states 
    under the FS maintains independent on $\alpha$ (and $\beta$) 
    in the limit of $K_F \rightarrow \infty$ whereas
    for finite $K_F$ it depends only slightly due to the effects
    on the edge of the FS.

Replacing $K_y$ by its actual value at the r-th channel (\ref{ky}) the 
Eq. (\ref{elliptical}) yields for the 1D radius $K_x$ of the 
FS in this channel

    \begin{equation}
    K_{x}=\alpha K_{F} \sqrt{1 - \left(\frac{k_{y}(r)}{\beta 
    K_{F}}\right)^{2}}.
    \end{equation}
The current in the whole 2D system is the sum of $M+1$ single channel 
currents (\ref{I1D}) taken with the appropriate $K_x$

    \begin{equation}
    \label{prad1}
    I_{M}(\phi)= \sum_{r=0}^{M} \sum_{l=0}^{\infty} \frac{2 I_{0}}{\pi 
    l}\alpha
    \sqrt{1 - \left( \frac{k_{y}(r)}{\beta K_{F}}\right)^{2}} \cos \left[l 
    K_{F}
    L_{x} \alpha \sqrt{1 - \left( \frac{k_{y}(r)}{\beta K_{F}}\right)^{2}}
    \right] \sin \left(\frac{2 \pi l\phi}{\phi_{0}}\right),
     \label{im1}
    \end{equation}
where $I_0=\frac{e\hbar K_{F}}{mL_{x}}$.

For the circular Fermi surface ($\alpha=\beta=1$) in a 44-channel system
for $L_{x}=1000 \AA$, $L_{y} = 100 \AA$ and assuming $E_{F}= \frac{\hbar^2
K_{F}^{2}}{2m}=7 eV$ (typical for Cu) the $I\!\!-\!\!\phi$ characteristic
is shown in Fig.1 \marginpar{insert Fig.1} (solid line). It is a ragged 
function, built of intervals
of the linear decrease and discontinuous jumps up of the current. Each
discontinuous increase of the current value in Fig.1 corresponds to a
single electron leaving or coming under the Fermi surface. This structure
of the current function is characteristic for $T=0$ and remains the same in
other cases in this chapter. 

If the parameter $\alpha$ increases then the circular FS turns into an
ellipse with the major axis along the $k_{x}$-axis.  An increase of $\alpha$
up to 10 (Fig.1 - dashed line) results in the decrease of the number of
channels ($M \sim \beta$) and the current value.  In this case we don't
observe any correlation between currents from different channels. 
   An increase of the parameter $\beta$ is equivalent to turning the major
axis of the ellipse along $k_{y}$-axis. If $\beta$ increases up to 10, then
a substantial enhancement of the current is observed (Fig.1 - dotted line). 
This is because the perpendicular to the $k_x$ axis part of the FS is 
very flat (its radius of curvature is proportional to 
$\frac{\beta}{\alpha}$) causing an enhancement of the current correlation.

It is worth noticing that the current correlation change similarly for 
the circular FS ($\alpha = \beta = 1$) when the circumference and the 
height of the cylinder are changed in such a way that $L_x L_y = const.$. 
In this case an increase of the cylinder height produces the same 
enhancement of current correlation as the increase of $\beta$. The total 
current however is even more enhanced because, at the same time, the 
circumference of the cylinder decreases causing the increase of $I_0$ (cf 
(\ref{im1})).

   \item[(ii)]  $K_F = const.$,$\alpha=\beta=1$, n is varied: 
    \label{dwa}  
    
    \begin{equation}
    K_{F}^{n}=  \left|K_{x}\right|^{n}+K_{y}^{n}.
    \end{equation}
      
    Here we consider $n$ to be an arbitrary positive real number. 
    For $n=1$ the Fermi surface is of a triangular
    shape (Fig.2)\marginpar{insert Fig.2}. With increasing n it becomes convex 
    and changes from 
    triangular 
    through circular $(n=2)$ to rectangular for $n \rightarrow \infty$. 
    On the other hand, decreasing n below $n=1$ yields  
    concave Fermi 
    surfaces shown in Fig.3 \marginpar{insert Fig.3}. Such FS are frequently 
    observed in HTSC materials. In our numerical
    calculations we adjust  $K_F$ to $n$ in such a way that the 
    volume under the FS remains constant.

     In the present case $K_{x}$ is

    \begin{equation}
    K_{x}=K_{F} \sqrt[n]{1 - \left(\frac{k_{y}(r)}{K_{F}}\right)^{n}}.
    \end{equation}
Inserting this into (\ref{ifi}) and (\ref{izero}) yields

    \begin{equation}
    I_{M}(\phi) = \sum_{r=0}^{M} \sum_{l=0}^{\infty} \frac{2 I_{0}}{\pi l}
    \sqrt[n]{1 - \left(\frac{k_{y}(r)}{K_{F}}\right)^{n}} \cos \left[l K_{F} 
    L_{x}\sqrt[n]{1 - \left(\frac{k_{y}(r)}{K_{F}}\right)^{n}}\right] \sin 
    \left(\frac{2 \pi l \phi}{\phi_{0}}\right).
    \end{equation}
 The case with  $n=0.4$ corresponds to the
FS from Fig.3. It is the case where the number of channels is 
greater compared to the circular FS ($n=2$). Taking again  $L_{x}=1000 
\AA$, $L_{y} = 100 \AA$ and  $E_{F}=7 eV$ the $I\!\!-\!
\!\phi$ characteristic for $n=0.4$ is shown in Fig.4 \marginpar{insert Fig.4}
(dotted line). We observe here a moderate 
increase of the current value compared to the circular FS (solid line). For 
$n = 1$ the FS is triangular and in the 
$I\!\!-\!\!\phi$ characteristic (Fig.4 - dashed line) there is a substantial
increase of correlation of phases from different
channels and the greater current value.
It is because all the states at the FS leave (or 
get in) the FS simultaneously. More generally, in case of $n=1$, there is 
a full correlation and enhancement of the current provided the following 
geometrical condition

\begin{equation}
  \frac{1}{2}\alpha L_x = \mu \beta L_y,
   \label{geom}
\end{equation}
where $\mu$ is an arbitrary positive integer, is obeyed. In case of the dashed 
line of Fig.4, $\mu=5$.

However, if the relation (\ref{geom}) do not hold the I($\phi$) current
can be much smaller. Therefore the geometrical amplification of the
current in case of $n=1$ is a matter of a careful adjustment of the cylinder
dimensions $L_x$ and $L_y$ in such a way that, for a given $\alpha$ and 
$\beta$, the condition (\ref{geom}) is fulfilled. Such a dependence on a
cylinder dimensions does not take place if we consider persistent currents
for the systems with $n>2$. For $n = 5$ we have a case in between circular
and rectangular FS. In the $I\!\!-\!\!\phi$ characteristic (Fig.5 \marginpar
{insert Fig.5}dotted line) some correlation among the channel currents exists 
what results in
an increase of the current amplitude compared to the circular FS (solid
line). The case with $n \rightarrow \infty$, i.e. rectangular FS, is the
most suitable to obtain a big current amplitude.  The $I\!\!-\!\!\phi$
characteristic is shown in Fig.5 (dashed line). We observe here a perfect
coherence (all the terminal states simultaneously leave or get into the
Fermi surface) and great increase of the current value. The amplitude of
the current is equal to the number of channels M in $I_{0}$-units. Thus in
general, with the departure of $n$ from 2, we observe the increasing 
amplitude of the M-channel current.

   \item[(iii)]  $\alpha\ll\beta$, $n=2$, $K_F$ is varied:
    \label{trzy}

In the previous two cases we have changed the shape of the FS leaving the
area under it invariant. As a result of that some currents were enhanced
and some were suppressed compared to the spherical FS. Changing parameters
$\alpha$, $\beta$ and $n$ we could not, however, predict whether the
current is positive or negative in a given interval. This important
property of the current is governed by the area under the FS. In Fig.6
\marginpar{insert Fig.6}
the currents are shown for an elliptical FS with $\beta \gg \alpha$.
Different current characteristic in this figure
correspond to $K_F$ increased up to 2\% from its initial value. The plot
shows that in this way the paramagnetic currents can be converted into
diamagnetic and vice versa. The $I\!\!-\!\!\phi$ characteristic is
therefore an almost periodic function of the volume under the FS. This
important property is reminiscent of the current behavior in a single
channel case, when the number of electrons in the system changes by two
\cite{Ho, szo1}. Here we show that this property of a single channel
survives the transition to the 2D case. 

 \end{description}

So far we considered a perfect cylinder. However it has been
shown \cite{ww,cheu,enti} that the phenomenon will survive modest
scattering, both elastic and inelastic. Let us assume that our mesoscopic
system contains a small number of
impurities (ballistic regime). The average current, where the
average is taken over impurity configurations, has been
calculated in \cite{enti}. The formula for the total average
current for 2D cylinder is
    \begin{equation}
    \label{avcu}
    \bar{I}(\phi)= I(\phi) \exp\left(-\frac{L_{x}}{2\lambda}\right), 
    \end{equation}
where $\lambda$ is the mean free path. For a sufficiently clean material
$\lambda$ is of the order of a few microns and in the mesoscopic regime the
impurities do not decreases the current significantly.

\section{Spontaneous Currents in the Mesoscopic Cylinder}
         
   In this chapter we generalize our considerations to 3D 
systems and to non zero temperatures. The formula (\ref{ifi}) for a current 
induced by the magnetic field in the 1D system of the mesoscopic 
size in $T>0$ is replaced by 
    \begin{equation}
    \label{it}
    I(\phi)=\sum_{l=1}^{\infty} \frac{4I_{1}T}{\pi T^{*}} \frac{\exp\left(-
    \frac{lT}{T^{*}}\right)}{1 - \exp\left(-\frac{2lT}{T^{*}}\right)} \cos
    (lK_{x}L_{x}) \sin \left(\frac{2 \pi l \phi}{\phi_{0}}\right),
    \end{equation}
where
$T^{*}$ is the characteristic temperature, defined by the energy gap
between energy levels at the Fermi surface

    \begin{equation}
    T^{*}= \frac{\hbar^{2}N}{mL_{x}^{2}}.
    \end{equation}
The formula (\ref{it}) is valid assuming the grand canonical 
ensemble \cite{Ho}. It is applicable in our case because single channels 
can exchange electrons according to $K_x$, which determines their 
chemical potential. Temperature influences the current in such a way, 
that all discontinuities in
the $I\!\!-\!\!\phi$ characteristic (all transitions in the current 
value) are smoothed, and the maximum of the amplitude decreases.

   Consider now a 3D system of the cylinder geometry  
made of a clean non superconducting material. The circumference, height and 
the width of the wall are denoted by $L_{x}, L_{y}$ and $L_{z}$ respectively. 
We assume that the width of the cylinder is small compared to the other 
dimensions $L_{z}\ll L_{x}, L_{z}\ll L_{y}$ and therefore we can assume with 
a good approximation that the vector potential {\bf A} does not depend on 
$z$.

Generalizing equation (\ref{dwym}) to a 3D system the 
current can be obtained by replacing $K_{x}$ in (\ref{it}) by
    \begin{equation}
    K_{x}=\alpha K_{F}\sqrt[n]{1 - 
    \left(\frac{k_{y}(r)}{\beta K_{F}}\right)^{n} - \left(\frac
    {k_{z}(s)}{\gamma K_{F}}\right)^{n}},
    \end{equation}
where $\alpha \beta \gamma = 1$ and
    \begin{equation}
    k_{z}(s)= \frac{s\pi}{L_{z}},\; \; \; \; s=0,1,2,\ldots.
    \end{equation}  
Therefore we obtain

    \begin{eqnarray}
    \label{ip}
    I(\phi) & = &
    \sum_{r=0}^{M}\sum_{s=0}^{P}\sum_{l=0}^{\infty}\frac{4I_{0}T\alpha}
    {\pi T^{*}} \sqrt[n]{1 - \left(\frac{k_{y}(r)}{\beta K_{F}}\right)^{n} - 
    \left(\frac
    {k_{z}(s)}{\gamma K_{F}}\right)^{n}} 
    \frac{\exp\left(-\frac{lT}{T^{*}}\right)}{1 -
    \exp\left(-\frac{2lT}{T^{*}}\right)} \times \nonumber \\
    & & \times \cos\left[lK_{F}L_{x} \alpha \sqrt[n]{1 - \left(\frac
    {k_{y}(r)}{\beta K_{F}}\right)^{n} - 
    \left(\frac{k_{z}(s)}{\gamma K_{F}}\right)^{n}}
    \right]\sin \left(\frac{2 \pi l \phi}{\phi_{0}}\right).
     \label{sc1}
    \end{eqnarray}
Changing the Fermi surface in the same way as in 2D case yields
a very similar effect. In cases where the current coherence was high the 
third dimension substantially increases 
the number of channels ($P \sim L_{z}$) and enhances the current. 

It has been recently shown in a series of papers \cite{Sz, szz} that a
metallic or a semiconducting system made of a set of mesoscopic
quasi 1D rings stacked along certain axis can
exhibit a transition to a low temperature state with a
spontaneous orbital current. 

In this chapter 
we concentrate on searching for a possibility of obtaining such spontaneous
currents in a clean metallic or semiconducting 3D cylinder. We consider here 
the case when the magnetic field $\phi$ enclosed by the cylinder, 
which drives the persistent current $I(\phi)$ around it, is the
sum of the externally applied flux $\phi_{e}$ and of the flux $\phi_{I}$
from the persistent current itself \cite{Sz}

    \begin{equation}
    \label{fi}
    \phi = \phi_{e} + {\cal L}I,
    \end{equation}
where $\cal L$ is the self-inductance of the cylinder (for cylinder with height
$L_{y}$, width $L_{z}$ and a radius $R=\frac{L_{x}}{2\pi}$, ${\cal L} = \frac{
\mu_{0} \pi R^{2}}{L_{y}^{2}} (\sqrt{L_{y}^{2}-R^{2}}-R)$). Most of
theoretical
discussions \cite{cheu, Ho} neglect the second term in (\ref{fi}).  Relations 
(\ref{fi}) and (\ref{ip}) represent the system of two selfconsistent equations 
to calculate the current 
$I(\phi)$ and the magnetic flux $\phi$. The question of existence of
the spontaneous persistent
currents reduces into a problem whether these equations have
stable, non-vanishing solutions at $\phi_{e}=0$

    \begin{equation}
    \label{os}
    \frac{\phi}{\cal L} = I(\phi).
     \label{sc2}
    \end{equation}
If there is any nonzero solution of (\ref{os}) then we have the spontaneous
current, corresponding to the point of intersection the straight line (LHS of
(\ref{os})) and the nonlinear curve (RHS of (\ref{os})) which is given by 
(\ref{ip}). The nonzero solution is stable provided the intersection is at the
place where the slope of the curve (\ref{ip}) is negative.

     We have made model calculations for a cylinder with the following
parameters $L_{x}= 25000 \pi \AA, L_{y}=10000 \AA, L_{z}=100 \AA, E_{F} =
7 eV$ with $\alpha=\beta=\gamma=1$ and $n=4,5$ (Figs 7, 8)\marginpar{insert 
Figs 7, 8}. We see that
for $n\geq4$ there is always a nonzero stable solution at sufficiently low
temperatures (circles at Figs 7 and 8). The temperature at which the 
spontaneous current occurs
increases with increasing $n$. One can show also that the condition
$\alpha=\beta=\gamma=1$ is not necessary and the spontaneous currents
occur even if the above parameters are different. It means that for all
the Fermi surfaces similar to rectangular the spontaneous currents can be
obtained. 

Similar transition to a spontaneous currents occur for a 
generalized triangular FS i.e. for $n=1$. In 
this case the FS is of a tetrahedral shape and the geometrical 
condition necessary to obtain spontaneous currents is

\begin{equation}
  \frac{1}{2}\alpha L_x = \mu \beta L_y =\nu\gamma L_z,
   \label{geom2}
\end{equation}
where $\mu$ and $\nu$ are positive integers.

   For 2D systems we do not observe spontaneous currents neither for high
value of n nor for reasonable values of $\alpha$ and $\beta$ coefficients.

\section{Conclusions}

   In this paper we presented same model considerations of persistent 
currents in clean multichannel systems of a cylindrical geometry (ballistic
regime).

   We discussed persistent current behavior due to the pure Aharonov-Bohm
effect --- we neglected attenuation of the current due to the field
penetrating through the system. We also ignored the spin-orbit coupling.
One can estimate that such assumptions \cite{Ho} pose no serious problems
for the cylinders under study. We also neglected here the Coulomb
interaction. Recently it has been shown that the Coulomb
interaction does not influence the persistent current behavior in clean 
systems \cite{avi}, whereas it enhances the current in the diffusive 
regime \cite{berk}. There is also an indication that the Coulomb 
interaction of the Hubbard type enhances the typical value of the 
persistent current in disordered rings in any dimensions \cite{rami}. The 
decrease of currents with disorder was taken into account (see eq. 
\ref{avcu}) following \cite{enti}. 

 The current in multichannel system depends strongly on the phase 
correlations between currents of different channels. 
We discussed persistent currents as a function of a shape of the Fermi
surface by comparing them to the standard circular FS. In case of the
elliptical FS the current-flux characteristics show substantial increase
of the current magnitude with increasing and attenuation with decreasing
curvature of the Fermi surface at the point where it crosses the $k_x$
axis. The triangular FS enhances the current provided its parameters obey 
the geometrical condition (\ref{geom}). The departure into concave FS 
only slightly increases the current. In the opposite case of the convex 
FS the current amplitude increases from its value in case of the circular 
FS to the maximal value in case of the rectangular FS ($n \rightarrow 
\infty$). We found also that the current-flux characteristic is an almost 
periodic function of an area under the FS.

   We also investigated the possibility of spontaneous currents in 2D and
3D cylinders. We found that for 3D cylinders spontaneous currents can be
obtained for $n\geq4$ and for $n=1$ provided the geometrical dimensions of
the cylinder are properly adjusted. This can be a hint for a possible
experimental verification of the currents. The characteristic temperatures
$T_{c}$ increase with increasing $n\geq4$. The most favorable situation
takes place for $n\rightarrow \infty$ what is equivalent to perfect
correlation of currents from different channels. Such Fermi surfaces can
be obtained e.g. for bcc crystals in tight binding approximation for
nearly half filled or half filled band. Preferable materials would be 
metals or semiconductors with a long phase coherence length and with a small 
number of impurities.

\section{Acknowledgments}

   Work was supported by Grant KBN PB 1108/P03/95/08. E.Z. is grateful to
Prof. W.Brenig and Prof. W.Zwerger for useful discussions. 

\newpage
\section{Figure captions}

Figure 1. Persistent current vs flux at T=0, for elliptical Fermi surfaces 
($n=2$), with $\beta=0.1$ (dashed line), $\beta=1$ (solid line) and 
$\beta=10$ (dotted line). In the inserted circle the magnified curve $(\alpha =
 \beta = 1)$ show the details of the current behavior. \vspace{0.5cm}\\
Figure 2. Convex Fermi surfaces according to Eq.(\ref{fersurf}) for 
$\alpha=\beta=1$ and $n=1,2,5$ and $\infty$.\vspace{0.5cm}\\
Figure 3. Concave Fermi surfaces according to Eq.(\ref{fersurf}) for
$\alpha=\beta=1$ and $n=0.8, 0.6$ and $0.4$.\vspace{0.5cm}\\
Figure 4. Persistent current vs flux at T=0 and $\alpha=\beta=1$, for 
spherical, $n=2$ (solid 
line), triangular, $n=1$ (dashed line) and concave $n=0.4$ (dotted line)
Fermi surfaces.\vspace{0.5cm}\\
Figure 5. Persistent current vs flux at T=0 and $\alpha=\beta=1$, for
convex Fermi surfaces, $n=2$ (solid line), $n=5$ (dotted line) and 
rectangular FS, $n\longrightarrow \infty$ (dashed line).\vspace{0.5cm}\\ 
Figure 6. Persistent current vs flux at T=0 and $\beta\gg\alpha$ and $n=2$, for
the elliptical Fermi surface and different volumes 
$K_{F_{1}}<K_{F_{2}}<\ldots<K_{F_{5}}$. The characteristic is an almost 
periodic function of $K_F$ (only half of the period in $K_{F}$ is 
shown). \vspace{0.5cm}\\
Figure 7. The graphical solution of a set of selfconsistent equations 
(\ref{sc1}) and (\ref{sc2}) for different temperatures and the convex 
Fermi surface $n=4$, $\alpha=\beta=1$. The nonzero crossings of the 
straight line (\ref{sc2}) with  the current-flux characteristic 
(\ref{sc1}) denoted by circles correspond to spontaneous currents in the 3D 
cylinder. \vspace{0.5cm}\\
Figure 8. The graphical solution of a set of selfconsistent equations 
(\ref{sc1}) and (\ref{sc2}) for different temperatures and the convex 
Fermi surface $n=5$, $\alpha=\beta=1$. The nonzero crossings of the 
straight line (\ref{sc2}) with  the current-flux characteristic 
(\ref{sc1}) denoted by circles correspond to spontaneous currents in the 3D 
cylinder.

\end{document}